\documentclass[aps,prb,twocolumn,showpacs,groupedaddress,preprintnumbers,amsmath,amssymb]{revtex4}
\usepackage[dvips]{graphicx}
\usepackage{bm}
\begin{document}

\title{Phenomenological theory of current driven exchange switching \\ in ferromagnetic nanojunctions}
\author{E.~M.~Epshtein,
Yu.~V.~Gulyaev, and P.~E.~Zilberman} \email{zil@ms.ire.rssi.ru}
\affiliation{Institute of Radio Engineering and Electronics of the
Russian Academy of Sciences, \\
Fryazino, Moscow Region, 141190, Russia}
\date{\today}
\begin{abstract}
Phenomenological approach is developed in the theory of spin-valve type
ferromagnetic junctions to describe exchange switching by current flowing
perpendicular to interfaces. Forward and backward current switching
effects are described and they may be principally different in nature.
Mobile electron spins are considered as free in all the contacting
ferromagnetic layers. Joint action of the following two current effects is
investigated: the nonequilibrium longitudinal spin-injection effective
field and the transverse spin-transfer surface torque. New vector boundary
conditions are derived, which represent, in essence, the continuity
condition for different spin fluxes. Dispersion relation for fluctuations is derived and solved for a
junction model having spatially localized spin transfer torque: depth of
the torque penetration into the free layer is assumed much smaller than
the total free layer thickness. Some critical value $\kappa_0$ of the well
known Gilbert damping constant $\kappa$ is established for the first time.
Spin transfer torque dominates in the instability threshold determination
for small enough constants $\kappa<\kappa_0$, while the spin-injection
effective field dominates for $\kappa>\kappa_0$. Typical estimation gives
$\kappa_0\sim 3\times 10^{-2}$ and it shows that the both mechanisms
mentioned may be responsible for the observable threshold in ferromagnetic
metal free layer used at room temperature. Fine interplay between spin
transfer torque and spin injection is necessary to provide a hysteretic
behavior of the resistance versus current dependence. The state diagrams
building up show the possibility of non-stationary (time dependent)
nonlinear states arising due to instability development. Calculations lead
to the instability increment values of the order of $\lesssim 10^{10}\,\rm
s^{-1}$ and they are really containing the contributions from the two
mechanisms mentioned. Spin wave resonance frequency spectrum softening
occurs under the current growing to the instability threshold.
Magnetization fluctuations above the threshold rise oscillatory with time
for small damping ($\kappa<\kappa_0$), but rise aperiodically and much more
rapid for large damping ($\kappa>\kappa_0$).
\end{abstract}
\maketitle

\section{Introduction}\label{section1}
Spin valve type magnetic junctions are of very interest for investigations
and subsequent applications. A number of significant effects were
predicted and experimentally revealed in the junctions last years.
Magnetization instability was predicted in Refs.~\onlinecite{Slonczewski,
Berger}. The instability is driven by the current perpendicular to plane
(CPP regime) due to near interface transformation of a transverse part of
mobile electron spin flux to the lattice magnetization flux. This
mechanism is usually referred to as a surface \emph{transverse spin
transfer} (TST) torque. Another mechanism was proposed and theory
elaborated in Refs.~\onlinecite{Heide,Gulyaev1,Gulyaev2,Gulyaev3}. This
second mechanism is due to nonequilibrium \emph{longitudinal spin
injection} (LSI) by current into the bulk of a ferromagnetic layer. The
injected spins create current dependent \emph{sd} exchange in nature
effective field, which acts on the lattice magnetization and may stimulate
reorientation phase transition. Theory of the joint action of the two
mechanisms mentioned was developed recently~\cite{Gulyaev4} (see also
preprints~\cite{Gulyaev5,Gulyaev7} and
Ref.~\onlinecite{Gulyaev9}).\footnote{Note that both mechanisms, TST
torque and LSI effective field, may act due to the same \emph{sd} exchange
interaction between mobile electron spins and bound electrons providing
the lattice magnetization,~\cite{Nagaev} but TST and LSI are the different
ways of the action. It will be treated further in more detail.}

Current driven instability was revealed experimentally, being manifested
in resistance jumps~\cite{Myers} and in microwave emission.~\cite{Tsoi}
Many interesting subsequent experiments confirmed the first observations
in principle and gave a lot of additional significant information (see,
e.g., Refs.~\onlinecite{Albert,Wegrowe,Grollier2,Chen,Huai,Krivorotov}).
The most of experiments considered trilayers Co/Cu/Co as a constitutive
working part of the junction. The first Co layer has usually pinned
lattice due to large its own thickness or existence of additional pinning
layers. The second Co layer is taken thin enough to have free lattice
magnetization, which may be controlled by external field or current. It is
significant to remark that thickness $L$ of the second free layer cannot
be done less than $\sim 2\,\rm{nm}$ to prevent a non-continuous structure
appearing.~\cite{Myers} Typically, the thicknesses in experiments were
chosen within the range $L\sim 2.5$--10 nm.

Success in experimental study and in applications stimulated an activity
to achieve more detailed understanding of the fundamental idea of TST
process introduced in Refs.~\onlinecite{Slonczewski, Berger}. Contrary to
this, in the present paper we accept this idea without any additional
proof: an abrupt transformation appears of transverse mobile electron spin
flux to the lattice magnetization flux at the interface between the
layers. Great attention was paid already to justify this idea more
rigorously starting from the microscopy theory approach (see, e.g.,
Refs.~\onlinecite{Stiles,Stiles2,Edwards,Edwards2,Zhang,Rebei}).
However, the phenomenological approach is necessary also because of its
more wide sight.

Exactly such an approach we laid in the basis of the present paper. In
other words, we accept the mobile electrons cannot perform long time and
long distance precession around the lattice magnetization because of phase
mismatch of individual electrons due to their statistical velocity spread.\cite{Slonczewski, Berger}
As a result, the transverse component of the mobile electron magnetization
vector decays rapidly and transfers to the lattice of ferromagnet. This
picture is taken as the base of our consideration and then all the
phenomenology is built up.

Such an approach allows us to integrate in the theory the second mechanism
of current action that is LSI. Therefore the LSI effective field and TST
torque are treated simultaneously as the two manifestations of the only
\emph{sd} exchange interaction. Additionally, we shall show such
significant effects as the backward current switching and the hysteretic
resistance versus current dependence may be understood as a fine interplay
of the two mechanisms mentioned. As far as we know, similar unified
description including the both mechanisms, TST and LSI, could not
developed yet in the frame of a microscopic theory. Moreover, we calculate
in the present paper not only the threshold current (that was done in many
papers) but the property of the various magnetic configurations: what
configurations remain stable and what ones become unstable depending on
the current. Similar experimental data were presented in the literature
before and we have now a possibility to compare some of them with the
theory. We show (similar to Refs.~\onlinecite{Gulyaev4,Gulyaev5}) the spin
wave spectrum of the junction layer should soften to zero under the
current is growing and tends to the instability threshold. That is
decisive manifestation of the LSI mechanism, which specifies it in
comparison with the TST one. We estimate also the rise time for the
unstable fluctuations. The minimal rise time may be less than $\sim
0.1\,\rm ns$ but only if the LSI mechanism of instability is dominated.

Let us make some more preliminary remarks. The first remark concerns to
the model of the system under consideration. According to the original
estimations~\cite{Slonczewski,Berger} and subsequent theory
calculations,~\cite{Stiles,Stiles2,Edwards,Zhang} the depth of transverse
spin flux penetration into the free layer is of the order of electron
quantum wavelength at the Fermi surface $\lambda_F\lesssim 1$--3 nm and is
strongly nonuniform within this interval. The spin transfer torque acts,
as it is well known,~\cite{Slonczewski,Berger} in that interval only.
Therefore the uniform approximation of the torque throughout the whole
thickness $L$ seems to be very approximate at least for large enough
$L\gtrsim 4$--10 nm. Because of the model of the torque, which is
uniformly spread in the free layer, is traditional one, it needs, at
least, to be justified. Recently~\cite{Gulyaev4,Gulyaev5,Gulyaev7} we
proposed another model for the calculations, namely, to consider spin
transfer torque as a boundary condition at the pinned/free layer
interface. In essence, this model is an opposite limiting case with
respect to the uniform torque model. The obvious advantage of our model is
in its complete non-sensitivity to details of the processes inside the
sublayer adjusted to the interface and having thickness $\lambda_F$. More
detailed comparison of these two models will be discussed in Appendix.

The second remark concerns to the external magnetic field influence. We
assume the junction may be placed in some external magnetic field
$\mathbf{H}$, which lies in the plane of junction interfaces. But this
field is supposed to be less in magnitude in comparison with the layer
magnetic anisotropy field $H_a$, that is $|\mathbf H|\equiv H<H_a$. Such a
supposition is accepted for simplicity because we want focus our attention
here to current (not to field) driven magnetization reversal processes.
Recent papers~\cite{Grollier,Slavin} show at large enough fields $H\ge
4\pi M\gg H_a$ the situation may become much more complicated.

\section{Model considered}\label{section2}
We will use further the most elementary model of a ferromagnetic metal
based on the well known concept of free \emph{s}-electron (quasiparticle)
ideal gas interacting with the lattice\footnote{We mean lattice of
magnetic ions taken in the continuum media approximation.} bound
\emph{d}-electron magnetization due to the so called \emph{sd} exchange
interaction.~\cite{Nagaev} Some more advanced (and more complicated!)
models may be used, of course. However, on our opinion, the most actual
problem now is to explain a number of experimental facts concerning to
switching of ferromagnetic nanojunctions. We believe it may be done in the
frame of the simple model.\footnote{We consider \emph{sd} interaction
similar for magnetic impurities and for proper magnetic ions of the
lattice. The difference might arise due to spatial distributions only:
random or periodic one. But all the distributions become identical in the
continuum approximation.}

\begin{figure*}
\includegraphics[scale=1.5 ]{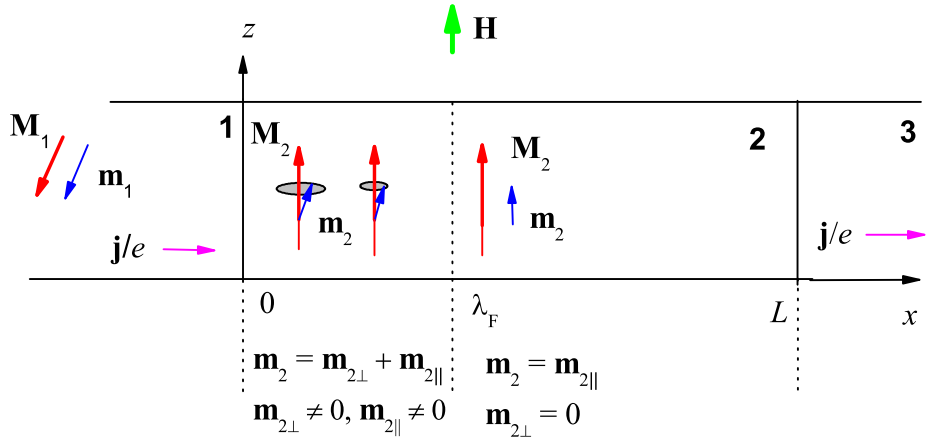}
\caption{(Color online) A scheme of magnetic junction to illustrate
processes in layer \textbf{2}. The contacting layers are labelled by
\textbf{1}, \textbf{2}, \textbf{3}. The arrows show directions of the
following vectors: $\mathbf{M}_1$ and $\mathbf{m}_1$ are magnetizations in
the layer \textbf{1}, $\mathbf{M}_2$ and $\mathbf{m}_2$ are magnetizations
in the layer \textbf{2}, external magnetic field $\mathbf{H}$ lies in the
junction plane $x=0$, and current density is $\mathbf{j}/e$ ($e$ is electron
charge). The vertical dashed lines separate two ranges in the layer
\textbf{2}. In $0\leq x\leq\lambda_F$ range, precession takes place, which
is shown with ovals. Vector $\mathbf{m}_2$ has both longitudinal and
transverse components $\mathbf{m}_{2\|}$ and $\mathbf{m}_{2\bot}$,
respectively. The precession angle decreases with $x$ increasing. In
$x>\lambda_F$ range, the precession vanishes, so that only the
longitudinal component remains.}\label{fig1}
\end{figure*}

We assume the junction consists of two ferromagnetic metal layers and one
very thin nonmagnetic spacer in between (shown as a heavy line in
Fig.~\ref{fig1}). To close the electric circuit, a nonmagnetic metal layer
\textbf{3} is necessary. The ferromagnetic metal layer \textbf{1} has
pinned orientation of the lattice magnetization $\mathbf{M}_1$, but the
mobile electron magnetization $\mathbf{m}_1$ is, in general, non-pinned.
Another ferromagnetic metal layer \textbf{2} has free lattice
magnetization $\mathbf{M}_2$ and mobile electron magnetization
$\mathbf{m}_2$, so that the magnetizations direction can be changed by an
external magnetic field $\mathbf H$ or spin-polarized current density
$\mathbf j$.

Plane $x=0$ is the interface between the layers \textbf{1} and \textbf{2}
and lattice magnetization $\mathbf{M}_1$ lies within the plane. Vectors
$\mathbf{M}_1$ and $\mathbf{M}_2$ make an angle $\chi$ of an arbitrary
value under the current is absent, that is under $\mathbf j=0$. After the
current is turned on, the electrons flow from the layer \textbf{1} into
the layer \textbf{2} or vise versa and then appear in a non-stationary
quantum state and ``walk'' between the spin energy subbands of the
corresponding layer. For forward current, when the inequality $j/e>0$ is
valid, vectors $\mathbf{m}_2$ and $\mathbf{M}_2$ perform a precession (see
Fig.~\ref{fig1}). As it was shown first in
Refs.~\onlinecite{Slonczewski,Berger}, angle of the precession decreases
with coordinate $x$ increasing and tends to zero at $x\approx\lambda_F$,
where $\lambda_F$ is an electron quantum wavelength at the Fermi surface.
This is because of electron velocity statistical spread and no relaxation
processes are needed to provide such a behavior. More detailed discussion
may be found also in the recent preprint.~\cite{Gulyaev5} Analogous
``precession'' region appears, of course, near the interface $x=0$ for
backward electron current, that is for $j/e<0$. However, the region will
be localized not in the layer \textbf{1} but again in the region
$0<x<\lambda_F$ of the layer \textbf{2}. It appears due to the complete
pinning of the lattice in the layer \textbf{1}. This feature is better to
discuss a little below in the paper.

The region $0<x<\lambda_F$ was introduced firstly by
Slonczewski~\cite{Slonczewski} and Berger~\cite{Berger} and further we
will refer to it as ``SB layer''. Their considerations of the layer were
based on the assumption that a ballistic transport occurs inside. An
opposite limiting case, when diffusion transport dominates, was considered
in Refs.~\onlinecite{Stiles,Zhang}. We follow, in this point, the approach
of the original works.~\cite{Slonczewski,Berger} The validity criterion
for this assumption may be written as $\lambda_F<l_p$, where momentum mean
free pass length $l_p$ may be estimated as $\sim 1$--10 nm for metals at
room temperature. As for the region $x>\lambda_F$  in the layer
\textbf{2}, the length $l_p$ appears indeed very small one. Based on this
reason, we will consider further the electron transport outside the SB
layer (in the depth of the layer \textbf{2}) using the diffusion--drift
limit.

\section{Lattice equations of motion}\label{section3}
Quantities that have no number indexes (1, 2 or 3) we refer further to any
of the junction layers. Motion of vector $\mathbf M$ will be described by
means of the Landau--Lifshitz--Gilbert (LLG) continuity equation:
\begin{equation}\label{1}
\frac{\partial\mathbf{M}} {\partial t}+\gamma
\left[\mathbf{M},\mathbf{H}_{eff}\right]-\frac{\kappa}
{M}\left[\mathbf{M},\frac{\partial\mathbf{M}} {\partial t}\right]=0,
\end{equation}
where $t$ is time, $\gamma$ is the gyromagnetic ratio, $\kappa$ is
dimensionless lattice damping constant ($0<\kappa\ll 1$),
\begin{equation}\label{2}
\mathbf{H}_{eff}=\mathbf{H}+\mathbf{H}_a+A\frac{\partial^2\mathbf{M}}
{\partial x^2}+\mathbf{H}_d+\mathbf{H}_{sd}
\end{equation}
is effective field, $\mathbf{H}_a=\beta(\mathbf{Mn})\mathbf n$ is
anisotropy field, $\beta$ being dimensionless anisotropy constant
(typically, $\beta\sim 0.1$ for our magnets) and $\mathbf n$ being a unit
vector along the anisotropy axis direction, $A$ is the intralattice
inhomogeneous exchange constant (magnetic ``stiffness'' of the lattice),
$\mathbf{H}_d$ is demagnetization field, $\mathbf{H}_{sd}$ is \emph{sd}
exchange effective field. The latter takes the form~\cite{Akhiezer}
\begin{equation}\label{3}
\mathbf{H}_{sd}\left(x\right)=-\frac{\delta U_{sd}}
{\delta\mathbf{M}\left(x\right)},
\end{equation}
where $\delta\left(\ldots\right)/\delta\mathbf{M}\left(x,\,t\right)$
 is a variational derivative, and $U_{sd}$ is \emph{sd} exchange energy,
\begin{eqnarray}\label{4}
U_{sd}=&-&\alpha_1\int\limits_{-\infty}^0\mathbf{m}_1\left(
x'\right)\mathbf{M}_1\,dx'\nonumber \\
&-&\alpha_2\int\limits_0^L
\mathbf{m}_2\left(x'\right)\mathbf{M}_2\left(x'\right)\,dx',
\end{eqnarray}
parameters $\alpha_1$ and $\alpha_2$ being dimensionless \emph{sd}
exchange constants (typical estimations are $\alpha_1\sim\alpha_2\sim
10^4$--$10^6$). Due to the last term in Eq.~\eqref{2}, the motion of
vectors $\mathbf{M}_2$, $\mathbf{m}_1$ and $\mathbf{m}_2$ appears to be
coupled. Remember, vector $\mathbf{M}_1$  is pinned completely in our
model and cannot be moved.

\section{Spatial distribution of mobile electron magnetization}\label{section4}
We describe the motion of vector $\mathbf m$  by means of continuity
equation
\begin{equation}\label{5}
  \frac{\partial\mathbf{m}}{\partial t}+\frac{\partial\mathbf{J}}{\partial
  x}+\gamma\alpha\left[\mathbf{m},\mathbf{M}\right]+\frac{\mathbf{m}-\bar{\mathbf{m}}}{\tau}=0,
\end{equation}
where $\tau$ is a time of relaxation to the local equilibrium value
$\bar{\mathbf m}\equiv\bar{m}\hat{\mathbf M}$, $\hat{\mathbf
M}\equiv\mathbf{M}/M$ is the unit vector, $\mathbf J$ is a mobile electron
magnetization flux density. We do not consider here the almost ballistic
quantum mechanical motion of the mobile electrons inside the SB layer.
Instead, we consider the influence of this small thickness layer on the
motion outside it by means of an appropriate boundary condition (see
below). Diffusion--drift approximation is applicable in the regions
outside SB layer, that is for $x>\lambda_F$. The equation~\eqref{5} may be
simplified significantly in this region.

First, only longitudinal part $\mathbf{m}_\|=m\hat{\mathbf M}$ may be
taken into account. Therefore an effective frequency $\omega$ of the
motion is determined by the precession in relatively small fields
$H,\,H_d,\,\beta M\ll\gamma\alpha
M\equiv\omega_{sd}\sim3\times10^{14}\,\rm s^{-1}$. We assume the condition
\begin{equation}\label{6}
  \omega\tau\ll 1
\end{equation}
is valid, which allows us to neglect time derivative in Eq.~\ref{5} in
comparison with the relaxation term. The remaining equation~\eqref{5} may
be solved then with respect to $\Delta\mathbf
m\equiv(\mathbf{m}-\bar{\mathbf m})=\Delta m\hat{\mathbf M}$. Let us take
the typical parameter estimations: $\alpha_1\sim\alpha_2\sim 2\times
10^4$, $M\sim 1\times10^3\,\mathrm G$, $\tau\sim 3\times 10^{-13}\,\mathrm
s$ to justify the following simplifying condition:
\begin{equation}\label{7}
  \omega_{sd}\tau\sim 10^2\gg 1.
\end{equation}
Then Eq.~\eqref{5} acquires the form
\begin{equation}\label{8}
  \Delta\mathbf m=-\tau\hat{\mathbf M}\left(\hat{\mathbf
  M}\cdot\partial\mathbf J/\partial x\right).
\end{equation}

Second, spin flux $\mathbf J$ can be found explicitly in the region
$x>\lambda_F$ to substitute it into Eq.~\eqref{8}. Electrons in the region
occupy energy subbands having their own spin magnetic moments parallel to
$\mathbf M\,\left(\uparrow\right)$ and antiparallel to $\mathbf
M\,\left(\downarrow\right)$. Magnetization of the mobile electrons
$\mathbf m$ contains contributions from the both subbands and therefore
may be presented as
\begin{equation}\label{9}
  \mathbf m=\mu_B\left(n_\uparrow-n_\downarrow\right)\hat{\mathbf M}\equiv
  m\hat{\mathbf M},
\end{equation}
where $\mu_B>0$ is Bohr magneton, $n_{\uparrow,\downarrow}$ are partial
electron densities in the spin subbands. The total electron density
$n=n_\uparrow+n_\downarrow$ does not depend on $x$ and $t$ because of
local neutrality conditions in metal. The magnetization flux density
$\mathbf J$ in Eq.~\eqref{8} also can be related with partial electric
current densities in each subbands, $j_\uparrow$ and $j_\downarrow$. The
relation has the following form:
\begin{equation}\label{10}
  \mathbf J=\frac{\mu_B}{e}\left(j_\uparrow-j_\downarrow\right)\hat{\mathbf M}.
\end{equation}
The total current density $j=j_\uparrow+j_\downarrow$ does not depend on
$x$ too, because of one-dimensional geometry of our model. We use standard
diffusion--drift formula for partial currents:
\begin{equation}\label{11}
  j_{\uparrow,\downarrow}=en_{\uparrow,\downarrow}\mu_{\uparrow,\downarrow}E
  -eD_{\uparrow,\downarrow}\frac{\partial n_{\uparrow,\downarrow}}{\partial x},
\end{equation}
where $\mu_{\uparrow,\downarrow}$ are partial mobilities,
$D_{\uparrow,\downarrow}$ are partial diffusion coefficients and $E$ is an
electric field strength inside the layer. Now, we should substitute
formula~\eqref{11} in Eq.~\eqref{10} and express $\mathbf J$ via $\mathbf
m$, using Eq.~\eqref{9}. The calculations are direct and were done
completely in Ref.~\onlinecite{Gulyaev2}. The result may be written as
follows:
\begin{equation}\label{12}
  \mathbf J=\left(\frac{\mu_B}{e}Qj
-\tilde{D}\frac{\partial m}{\partial x}\right)\hat{\mathbf M},
\end{equation}
where
$Q=\left(\sigma_\uparrow-\sigma_\downarrow\right)/\left(\sigma_\uparrow
+\sigma_\downarrow\right)$  may be understood as an equilibrium
conductivity spin polarization parameter with
$\sigma_{\uparrow,\downarrow}\equiv en_{\uparrow,\downarrow}\mu_{\uparrow,
\downarrow}$, and $\tilde D=\left(\sigma_\uparrow D_\downarrow
+\sigma_\downarrow
D_\uparrow\right)/\left(\sigma_\uparrow+\sigma_\downarrow\right)$ is the
effective spin diffusion constant. To obtain Eq.~\eqref{12}, an additional
assumption should be made,~\cite{Gulyaev2} namely
\begin{equation}\label{13}
  \frac{j}{j_D}\ll 1,
\end{equation}
where $j_D\equiv enl/\tau$  is a characteristic current density in the
layer \textbf{2}. With typical parameter values, $n\sim
10^{22}\,\mathrm{cm}^{-3}$, $l\sim 3\times 10^{-6}\,\mathrm{cm}$,
$\tau\sim 3\times 10^{-13}\,\mathrm s$, we get $j_D\sim 1.6\times
10^{10}\,\mathrm{A/cm}^2$. The condition means the current disturbs the
subband populations rather slightly, so that only a low injection level is
realized. As it was mentioned above, we are interesting in currents $j\le
10^7$--$10^8\,\mathrm{A/cm}^2$, which have an order of instability
thresholds. Therefore, the condition~\eqref{13} should be well satisfied
in our calculations.

After substituting expression~\eqref{12} into Eq.~\eqref{8} we obtain
finally the following equation for $m$:
\begin{equation}\label{14}
  \frac{\partial^2m}{\partial x^2}-\frac{\Delta m}{l^2}=0,
\end{equation}
where $\Delta m=m-\bar m$. Since the \emph{sd} exchange gap is fixed in
our model, $\bar m$ value does not depend on $x$ and $t$. The spin
diffusion length is $l=\sqrt{\tilde D\tau}$.

As it was mentioned at the beginning of this section, our above presented
derivation may be referred to any junction layer. Therefore,
Eq.~\eqref{14} is valid to describe any definite layer if we perform the
replacements: $m\rightarrow m_{1,2,3}$, $\hat{\mathbf
M}\rightarrow\hat{\mathbf M}_{1,2,3}$, $Q\rightarrow Q_{1,2,3}$, $\tilde
D\rightarrow\tilde D_{1,2,3}$, $\tau\rightarrow\tau_{1,2,3}$, and so on.
Moreover, we should take $\hat{\mathbf M}_3\equiv\hat{\mathbf M}_2$,
$\bar{m}_3=0$, $Q_3=0$ for nonmagnetic layer \textbf{3} (see
Ref.~\onlinecite{Gulyaev2} for more details).

To solve Eq.~\eqref{14}, we should derive the appropriate boundary
conditions. We accept two sorts of the conditions: 1) the continuity
condition of the mobile electron longitudinal spin fluxes, penetrating
through the interfaces and 2) the continuity condition for difference
between the chemical potentials of electrons in the spin subbands. It
means the following equalities are to be valid for current flowing in
forward direction ($j/e>0$): $\mathbf J_1(-\lambda_F)\hat{\mathbf
M}_2(\lambda_F)= \mathbf J_2(\lambda_F)\hat{\mathbf M}_2(\lambda_F)$ and
$\mathbf J_2(L-\varepsilon)\hat{\mathbf M}_2(L-\varepsilon)= \mathbf
J_3(L+\varepsilon)\hat{\mathbf M}_2(L-\varepsilon)$, as well the following
equalities are to be valid for backward direction of current ($j/e<0$):
$\mathbf J_2(\lambda_F)\hat{\mathbf M}_1= \mathbf
J_1(-\lambda_F)\hat{\mathbf M}_1$ and $\mathbf
J_3(L+\varepsilon)\hat{\mathbf M}_2(L-\varepsilon)= \mathbf
J_2(L-\varepsilon)\hat{\mathbf M}_2(L-\varepsilon)$. It may be written in
an explicit form using the definition \eqref{12}:
\begin{eqnarray}\label{15}
  &&\left(\frac{\mu_B}{e}Q_1j-\tilde{D}_1\frac{\partial m_1}{\partial
  x}\Bigl|_{x=-\lambda_F}\right)\left(\hat{\mathbf{M}}_1\hat{\mathbf{M}}_2(\lambda_F)\right)\nonumber \\
  &&=\frac{\mu_B}{e}Q_2j
  -\tilde{D}_2\frac{\partial m_2}{\partial x}\Bigr|_{x=\lambda_F}, \nonumber \\
  &&\frac{\mu_B}{e}Q_2j-\tilde{D}_2\frac{\partial m_2}{\partial
  x}\Bigl|_{x=L-\varepsilon}=-\tilde{D}_3\frac{\partial m_3}{\partial x}\Bigr|_{x=L+\varepsilon }
\end{eqnarray}
for $j/e>0$ and
\begin{eqnarray}\label{16}
  &&\left(\frac{\mu_B}{e}Q_2j-\tilde{D}_2\frac{\partial m_2}{\partial
  x}\Bigl|_{x=\lambda_F}\right)\left(\hat{\mathbf{M}}_1\hat{\mathbf{M}}_2(\lambda_F)\right)\nonumber \\
  &&=\frac{\mu_B}{e}Q_1j
  -\tilde{D}_1\frac{\partial m_1}{\partial x}\Bigr|_{x=-\lambda_F}, \nonumber \\
  &&-\tilde{D}_3\frac{\partial m_3}{\partial x}\Bigl|_{x=L+\varepsilon}=\frac{\mu_B}{e}Q_2j-\tilde{D}_2\frac{\partial m_2}{\partial
  x}\Bigr|_{x=L-\varepsilon}
\end{eqnarray}
for $j/e<0$.

The continuity condition for chemical potential differences was considered
before us but, apparently, only for collinear orientations of
magnetizations on the two opposite sides of the boundary
(spacer~\cite{Valet} or domain wall~\cite{Zvezdin}). We have written the
condition for oblique magnetization orientations in
Ref.~\onlinecite{Gulyaev8}. Now, we should return to the situation of
oblique magnetizations because: 1) the idealization of pinned mobile
electrons is removed and 2) vectors $\mathbf M_1$ and $\mathbf M_2$ may
appear noncollinear. After calculations analogous to
Ref.~\onlinecite{Gulyaev8}, we obtain
\begin{eqnarray}\label{17}
  &&N_1\Delta m_1(-\lambda_F)=N_2\Delta
  m_2(\lambda_F)\left(\hat{\mathbf{M}}_1\hat{\mathbf{M}}_2(\lambda_F)\right),\nonumber
  \\ &&N_2\Delta m_2(L-\varepsilon)=N_3\Delta m_3(L+\varepsilon)
\end{eqnarray}
\\for $j/e>0$ and
\begin{eqnarray}\label{18}
&&N_1\Delta m_1(-\lambda_F)\left(\hat{\mathbf M}_1\hat{\mathbf
M}_2(\lambda_F)\right)=N_2\Delta m_2(\lambda_F),\nonumber
\\ &&N_3\Delta m_3(L+\varepsilon)=N_2\Delta m_2(L-\varepsilon)
\end{eqnarray}
for $j/e<0$. Here the following definitions are introduced:
$N=(2\mu_B)^{-1}\left[(g_\uparrow (\bar\zeta))^{-1}+(g_\downarrow
(\bar\zeta))^{-1}\right]$, the quantities $g_{\uparrow,\downarrow}$ are
densities of states for spin up and spin down electrons, $\bar\zeta$ is
equilibrium chemical potential, which is constant across each junction
layer.

The equation~\eqref{14} may be directly solved with the following result:
\begin{eqnarray}\label{19}
&&\Delta m_1(x)=\tilde m_1\exp\left(\frac{x}{l_1}\right),\nonumber \\
&&\Delta m_2(x)=\tilde m_2\cosh\left(\frac{x}{l_2}\right)
+l_2\tilde m'_2\sinh\left(\frac{x}{l_2}\right),\nonumber \\
&&\Delta m_3(x)=\tilde m_3\exp\left(\frac{L-x}{l_3}\right),
\end{eqnarray}
where four constants $\tilde m_1$, $\tilde m_2$, $\tilde m'_2$, $\tilde
m_3$ should be found from the boundary conditions~\eqref{15}--\eqref{18}.
Moreover, we take $\Delta m_1\rightarrow 0$, $\Delta m_3\rightarrow 0$ far
from the interfaces, that is for $|x|\rightarrow\infty$. Then we obtain
the distribution of mobile electron magnetization across the junction
considered. After some direct calculations, we obtain the following
results for the forward current $j/e>0$:
\begin{widetext}
\begin{eqnarray}\label{20}
&&\tilde m_2=\mu_Bn_2\frac{j}{j_{D2}}\left\{Q_2+
\left[Q_1\left(\hat{\mathbf{M}}_1\hat{\mathbf{M}}_2(\lambda_F)\right)
-Q_2\right]\left(\cosh\lambda+\nu_2\sinh\lambda\right)\right\}\nonumber \\
&&\times\left[\sinh\lambda+\nu_2\cosh\lambda+
\frac{1}{\nu_1}\left(\hat{\mathbf{M}}_1\hat{\mathbf{M}}_2(\lambda_F)\right)^2\left(\cosh\lambda
+\nu_2\sinh\lambda\right)\right]^{-1},
\end{eqnarray}
\begin{equation}\label{21}
  l_2\tilde m'_2=\frac{1}{\nu_1}\left(\hat{\mathbf{M}}_1\hat{\mathbf{M}}_2(\lambda_F)\right)^2\tilde m_2
  -\mu_Bn_2\frac{j}{j_{D2}}\left[Q_1\left(\hat{\mathbf{M}}_1\hat{\mathbf{M}}_2(\lambda_F)\right)
  -Q_2\right],
\end{equation}
\begin{equation}\label{22}
  \tilde m_1=\frac{N_2}{N_1}\left(\hat{\mathbf{M}}_1\hat{\mathbf{M}}_2(\lambda_F)\right)\tilde
  m_2.
\end{equation}
\end{widetext}
Here we introduce dimensionless thickness $\lambda=L/l_2$ and matching
parameters for different interfaces $\nu_1=\tilde D_2l_1N_1/\tilde
D_1l_2N_2$, $\nu_2=\tilde D_3l_2N_2/\tilde D_2l_3N_3$.

Matching parameters are, in essence, the ratios of corresponding spin
contact resistances. For small parameters ($\nu_{1,2}\ll 1$) spin flux
penetrates slowly through the corresponding interface, that is the
injection weakens greatly. Similar situation is typical for
metal--semiconductor boundary (see, e.g., Ref.~\onlinecite{Fert}).
Opposite case of large enough parameter ($\nu_{1,2}\gg 1$) corresponds to
ready flux penetration through the interface, and the moderate penetration
corresponds to $\nu_{1,2}\approx 1$.

As it is seen clearly from formulae~\eqref{19}--\eqref{22}, the current
disturbs spin equilibrium deeply inside the junction layers at distances
of the order of diffusion lengths $l_{1,2,3}$. Therefore the
nonequilibrium spin injection by current is indeed a bulk effect.
Moreover, the spin injection\footnote{The term ``spin injection'' for a
disturbance of spin equilibrium by light (see, e.g.,
Ref.~\onlinecite{Zakharchenya}) or current is a traditional one starting from the
first papers~\cite{Aronov,Aronov2} and it corresponds closely to the
famous term ``charge injection'' in semiconductor
physics.~\cite{Shockley}} depends on the angle between the vectors
$\hat{\mathbf{M}}_1$ and $\hat{\mathbf{M}}_2(\lambda_F)$. It shows
explicitly the mobile electron magnetization does depend on the lattice
magnetization vector $\hat{\mathbf{M}}_2$  direction. In principal, such
dependence arises because the mobile electrons represent much more rapid
system than the magnetic lattice. Therefore the electrons ``see'' at any
time the instantaneous lattice configuration and are adapted to it.
Formally, it is seen from our transformation~\eqref{8} of the basic
equation~\eqref{5} for electrons. The transformation is based on the
condition~\eqref{6}. The latter condition allows neglecting any inertia in
the electron following the lattice.

Note finally according to boundary conditions~\eqref{15}--\eqref{18}, we
may simply make a replacement
\begin{equation}\label{23}
\left(\hat{\mathbf{M}}_1\hat{\mathbf{M}}_2(\lambda_F)\right)
\rightarrow\left(\hat{\mathbf{M}}_1\hat{\mathbf{M}}_2(\lambda_F)\right)^{-1}
\end{equation}
to go from the forward to backward current direction. Therefore, formulae
~\eqref{19} and~\eqref{20}--\eqref{22} remain valid for backward current
($j/e<0$) after the replacement~\eqref{23} is done.

\section{Calculation of current dependent \emph{sd} exchange effective field
$\mathbf{H}_{sd}(j,\,x)$}\label{section5}
We should calculate variational
derivative~\eqref{3} to find the field $\mathbf{H}_{sd}(j,\,x)$. First, we
calculate the \emph{sd} exchange energy substituting
formulae~\eqref{19}--\eqref{22} into expression~\eqref{4} and performing
the integration over coordinate $x$. Then, we obtain for the forward
current ($j/e>0$)
\begin{widetext}
\begin{eqnarray}\label{24}
  &&U_{sd}=U_{sd}^{(eq)}-\mu_B\alpha_2n_2M_2l_2Q_1\frac{j}{j_{D2}}\left[\sinh\lambda
  +\nu_2\cosh\lambda+\frac{1}{\nu_1}\left(\hat{\mathbf{M}}_1\hat{\mathbf{M}}_2(\lambda_F)\right)^2
  (\cosh\lambda+\nu_2\sinh\lambda)\right]^{-1}\nonumber \\
  &&\times\Biggl\{\frac{Q_2}{Q_1}\left[\sinh\lambda+\frac{1}{\nu_1}(\cosh\lambda-1)
  \left(\hat{\mathbf{M}}_1\hat{\mathbf{M}}_2(\lambda_F)\right)^2\right]+\left[\sinh\lambda
  +\nu_2(\cosh\lambda-1)\right]\left[\left(\hat{\mathbf{M}}_1\hat{\mathbf{M}}_2(\lambda_F)\right)
  -\frac{Q_2}{Q_1}\right]\nonumber \\
  &&+\frac{b}{\nu_1}\left(\hat{\mathbf{M}}_1\hat{\mathbf{M}}_2(\lambda_F)\right)
  \left[\frac{Q_2}{Q_1}+\left[\left(\hat{\mathbf{M}}_1\hat{\mathbf{M}}_2(\lambda_F)\right)
  -\frac{Q_2}{Q_1}\right](\cosh\lambda+\nu_2\sinh\lambda)\right]\Biggr\},
\end{eqnarray}
\end{widetext}
where $U_{sd}^{(eq)}$ is an equilibrium part of the energy, which depends
only on modulus $M_1$ and $M_2$. Parameter
$b=\alpha_1\tau_1M_1/\alpha_2\tau_2M_2$ describes the contribution of the
layer \textbf{1} into the energy~\eqref{4}. Note, that the exchange energy
for backward current ($j/e<0$) may be easily got from
expression~\eqref{24} by means of the replacement~\eqref{23}.

After differentiating~\eqref{24} in accordance with~\eqref{3}, the
equilibrium part of the effective field is collinear with $\mathbf M$ and
does not contribute to LLG equation~\eqref{1}. Nonequilibrium part of the
field may be written as
\begin{eqnarray}\label{25}
  \Delta\mathbf{H}_{sd}&=&-\frac{\partial U_{sd}}{\partial\left(\hat{\mathbf{M}}_1\hat{\mathbf{M}}_2
  (\lambda_F)\right)}\cdot\frac{\delta\left(\hat{\mathbf{M}}_1\hat{\mathbf{M}}_2(\lambda_F)\right)}{\delta
  \mathbf{M}_2(x)}\nonumber \\
  &=&h_{sd}\hat{\mathbf{M}}_1l_2\delta(x-\lambda_F),
\end{eqnarray}
where explicit expressions for the field $h_{sd}$ depend on results of
differentiation of the energy being different for forward and backward
currents. The most general expressions may be directly obtained but appear
too cumbersome. We show therefore some simple partial cases only:

a) forward current ($j/e>0$) and a very thin layer \textbf{2} ($\lambda\ll
1$)
\begin{eqnarray}\label{26}
  h_{sd}&=&\mu_B\alpha_2n_2Q_1\frac{j}{j_{D2}}\left[\nu
  +\left(\hat{\mathbf{M}}_1\hat{\mathbf{M}}_2(\lambda_F)\right)^2\right]^{-2}\nonumber \\
  &\times&\Biggl\{\lambda\nu_1\left[\nu
  -\left(\hat{\mathbf{M}}_1\hat{\mathbf{M}}_2(\lambda_F)\right)^2\right]\nonumber
  \\  &+&2b\nu\left(\hat{\mathbf{M}}_1\hat{\mathbf{M}}_2(\lambda_F)\right)\Biggr\},
\end{eqnarray}
where $\nu=\nu_1\nu_2=\tilde D_3l_1N_1/\tilde D_1l_3N_3$;

b) backward current ($j/e<0$) and $\lambda\ll 1$
\begin{eqnarray}\label{27}
  h_{sd}&=&\mu_B\alpha_2n_2Q_1\frac{j}{j_{D2}}\left[1
  +\nu\left(\hat{\mathbf{M}}_1\hat{\mathbf{M}}_2(\lambda_F)\right)^2\right]^{-2}\nonumber \\
  &\times&\Biggl\{\lambda\nu_1\left[1
  -\nu\left(\hat{\mathbf{M}}_1\hat{\mathbf{M}}_2(\lambda_F)\right)^2\right]\nonumber
  \\ &-&2b\nu\left(\hat{\mathbf{M}}_1\hat{\mathbf{M}}_2(\lambda_F)\right)\Biggr\}.
\end{eqnarray}

It is interesting to note that the general expression~\eqref{25} for
effective field is proportional to singular function
$\delta(x-\lambda_F)$. It means the field vanishes everywhere except the
SB layer near the interface $x=0$ taken from the side of the layer
\textbf{2}. Therefore the magnitude of the effective field is a bulk
quantity depending on such bulk parameters as $L,\,l_2,\,n_2,\,\nu$. But
this field is applied near to interface only, namely, to the point
$x=\lambda_F$.

This feature has a significant consequence. The action of the field may be
described by two equivalent ways. One of them is to consider
field~\eqref{25} as a singular term in the bulk equations of motion LLG
for the lattice magnetization. Exactly such a way was chosen in our
previous preprint~\cite{Gulyaev5} and article.~\cite{Gulyaev4} But much
more simple way is to include the field into the boundary condition for
the lattice magnetization at the boundary point $x=\lambda_F$. This last
way was proposed in Ref.~\onlinecite{Gulyaev7} and we develop it here in
the next section.

\section{Boundary conditions for lattice magnetization}\label{section6}
We suggest a model, which is opposite to the model of uniformly spread
spin transfer torque in the layer \textbf{2}. In our model, the spin
transfer torque is localized inside the SB layer $0<x<\lambda_F$ and may
be described as a boundary condition. Such a model corresponds to the real
situation if only the thickness of SB layer $\lambda_F$ occurs the
smallest length among the other lengths in the system (i.e.,
$L\gg\lambda_F$, $\sqrt A\gg\lambda_F$, and $l\gg\lambda_F$).

The total magnetization flux in the layers \textbf{1} and \textbf{3}
consists only of mobile electron fluxes because magnetic lattice in the
layer \textbf{1} is pinned, while the lattice in the layer \textbf{3} is
nonmagnetic one. Therefore, fluxes in these layers may be described by
means of expressions of the type~\eqref{12}. The situation with the layer
\textbf{2} occurs significantly more complicated: mobile electrons and
lattice both contribute to the flux. Mobile electrons contribute to
longitudinal part of the flux, which coincides with Eq.~\eqref{12}, while
the lattice creates a transverse part of the flux.

To derive all the spin fluxes, which may flow in our junction, it is
convenient to return to initial equations~\eqref{1} and~\eqref{5}, sum
them and rewrite in the form
\begin{eqnarray}\label{28}
  &&\frac{\partial(\mathbf M+\mathbf m)}{\partial t}+\gamma A\left[\mathbf
M,\,\frac{\partial^2\mathbf M}{\partial x^2}\right]+\gamma\left[\mathbf
M,\,\mathbf{H}_{sd}\right]+\frac{\partial\mathbf J}{\partial x}\nonumber \\
&&+\gamma\left[\mathbf M,\,\mathbf H'\right]+\gamma\alpha[\mathbf
m,\,\mathbf M] -\frac{\kappa}{M}\left[\mathbf M,\,\frac{\partial\mathbf
M}{\partial
t}\right]\nonumber \\
&&+\frac{\mathbf m-\bar{\mathbf m}}{\tau}=0,
\end{eqnarray}
where $\mathbf H'=\mathbf H+\mathbf H_a+\mathbf H_d$. The summation of the
initial Eqs.~\eqref{1} and ~\eqref{5} is necessary to do in our case
because the mobile electron flux may transfer to the lattice one and these
fluxes together determine dynamics of the system. The second summand in
the left hand side of Eq.~\eqref{28} may be rewritten as follows:
\begin{equation}\label{29}
  \gamma A\left[\mathbf M,\,\frac{\partial^2\mathbf M}{\partial^2 x}\right]
  =a\frac{\partial}{\partial x}\left[\hat{\mathbf M},\,\frac{\partial\mathbf M}{\partial x}\right]
  \equiv\frac{\partial\mathbf{J}_M}{\partial x},
\end{equation}
where
\begin{equation}\label{30}
  \mathbf{J}_M=a\left[\hat{\mathbf M},\,\frac{\partial\mathbf M}{\partial
  x}\right]
\end{equation}
may be considered as a lattice magnetization flux with parameter $a=\gamma
AM$  having the sense of a lattice magnetization diffusion constant (see
Refs.~\onlinecite{Gulyaev4,Gulyaev5} for more details).

The next (third) summand in Eq.~\eqref{28} may be transformed using
formula~\eqref{25} for $\mathbf{H}_{sd}$ to the following form:
\begin{equation}\label{31}
  \gamma\left[\mathbf M,\,\mathbf{H}_{sd}\right]=\frac{\partial\mathbf{J}_{sd}}{\partial
  x},
\end{equation}
where
\begin{equation}\label{32}
  \mathbf{J}_{sd}(x)=\gamma
  h_{sd}l_2\left[\mathbf{M}_2(\lambda_F),\,\hat{\mathbf{M}}_1\right]\theta(x-\lambda_F)
\end{equation}
represents a very interesting new quantity that may be called as an
``effective field transverse flux'' and $\theta(x-\lambda_F)$  is a well
known step function: $\theta(x-\lambda_F)=0$ for $x<\lambda_F $ and
$\theta(x-\lambda_F)=1$ for $x>\lambda_F $. Therefore, two lattice
magnetization fluxes exist: \eqref{30} and~\eqref{32}, the last appearing
due to \emph{sd} effective field. All the lattice fluxes are perpendicular
to the vector $\mathbf M_2$. The fourth summand in Eq.~\eqref{28}
originates from mobile electron spin flux $\mathbf J$, which is given by
Eq.~\eqref{12}. As it is seen, the flux $\mathbf J$ is directed along
vector $\mathbf M$.

\begin{figure}
\includegraphics[width=8cm]{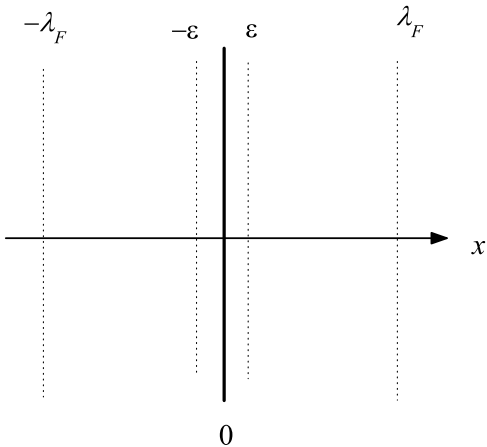}
\caption{Range of integration~\eqref{33}: the vertical dotted lines show
limits of integration.}\label{fig2}
\end{figure}

Let us introduce the integration over the near-interface regions
$-\lambda_F<x<-\varepsilon$ and $\varepsilon<x<\lambda_F$ with
infinitesimal parameter $\varepsilon\rightarrow +0$ (see Fig.~\ref{fig2}
for illustration):
\begin{equation}\label{33}
  \langle\ldots\rangle_-\equiv\int\limits_{-\lambda_F}^{-\varepsilon}(\ldots)\,dx
  \quad\text{and}\quad\langle\ldots\rangle_+\equiv\int\limits_\varepsilon^{\lambda_F}(\ldots)\,dx
\end{equation}
and apply these operations to Eq.~\eqref{28}. We should now introduce the
conditions, which say explicitly that our nonmagnetic spacer layer appears
nontransparent for the lattice magnetization excitations and fluxes. These
conditions may be written as
\begin{equation}\label{34}
  \mathbf J_M(-\varepsilon)=0,
  \quad\mathbf J_M(\varepsilon)=0,\quad\mathbf J_{sd}(\varepsilon)=0,
\end{equation}
the latter condition being satisfied automatically because of the
definition~\eqref{32}. Moreover, we accept the condition that our spacer
is completely transparent for mobile electron spin flux that is no spin
scattering processes occurs inside. This last condition may be written as
\begin{equation}\label{35}
  \mathbf J(-\varepsilon)=\mathbf J(\varepsilon).
\end{equation}

After the operations~\eqref{33} are applied, a number of simplifications
become possible. In particular, two last summands in the left hand side of
Eq.~\eqref{28} may be considered as a very small in comparison with the
first summand (since $0<\kappa\ll 1,\;\omega\tau\ll 1$). These relaxation
summands should be omitted if we want the flux continuity equations may be
valid approximately. Further, there exist a number of ``volume'' summands,
which are proportional to the small length $\lambda_F$. It is desirable
these summands to be sufficiently small too, namely,
\begin{eqnarray}\label{36}
  &&\lambda_F\left|\frac{\partial(\mathbf M+\mathbf m)}{\partial
  t}+\gamma\left[\mathbf M,\,\mathbf H'\right]+\gamma\alpha[\mathbf
  m,\,\mathbf M]\right|_{-\lambda_F<x<-\varepsilon}\nonumber \\
  &&\ll\left|\mathbf
  J(-\lambda_F)\right|,
\end{eqnarray}
\begin{eqnarray}\label{37}
  &&\lambda_F\left|\frac{\partial(\mathbf M+\mathbf m)}{\partial
  t}+\gamma\left[\mathbf M,\,\mathbf H'\right]+\gamma\alpha[\mathbf
  m,\,\mathbf M]\right|_{\varepsilon<x<\lambda_F}\nonumber \\
  &&\ll\left|\mathbf
  J(\lambda_F)+\mathbf{J}_{sd}(\lambda_F)\right|.
\end{eqnarray}
The conditions~\eqref{36} and~\eqref{37} characterize an applicability of
our model: the better they are satisfied, the more exact is the model.
Formally, conditions~\eqref{36} and~\eqref{37} are satisfied well in the
limit $\lambda_F\rightarrow 0$.

Then, we apply~\eqref{33} to~\eqref{28} using Eqs.~\eqref{34}, \eqref{35}
and making all the simplifications. We obtain as a result
\begin{equation}\label{38}
  -\mathbf J_M(-\lambda_F)- \mathbf J(-\lambda_F)+ \mathbf J(\varepsilon)=0,
\end{equation}
\begin{equation}\label{39}
  \mathbf J(\lambda_F)-\mathbf J(\varepsilon)+\mathbf J_M(\lambda_F)+\mathbf J_{sd}(\lambda_F)=0.
\end{equation}
We put $\mathbf{J}_M(-\lambda_F)=0$ in Eq.~\eqref{38} because of
$\mathbf{M}_1$ magnetization is pinned and, as we assume, cannot be
excited by the electric current. The conditions~\eqref{38} and~\eqref{39}
are valid for all the fluxes at the interface $x=0$. Analogously, we may
derive the conditions for the second interface $x=L$. The result is
\begin{equation}\label{40}
  \mathbf J(L+\varepsilon)-\mathbf J(L-\varepsilon)-\mathbf J_M(L-\varepsilon)=0,
\end{equation}
where we take into account the flux $\mathbf J_{sd}$ is constant at $x=L$
and therefore falls out from the continuity condition. The
conditions~\eqref{38}--\eqref{40} represent in the most general vector
form the quest boundary conditions. They allow us further to find all the
necessary solutions of our initial equations of motion~\eqref{1}
and~\eqref{5}.

In particular, we obtain immediately the longitudinal boundary
conditions~\eqref{15} and~\eqref{16} from Eqs.~\eqref{38}--\eqref{40}. To
show this for forward current $j/e>0$ we should simply multiply scalar
Eqs.~\eqref{38}--\eqref{40} by $\hat{\mathbf M}_2$. For $j/e<0$, we
multiply Eq.~\eqref{40} by $\hat{\mathbf M}_2$ and multiply
Eqs.~\eqref{38} and~\eqref{40} by $\hat{\mathbf M}_1$. Note, after the
multiplication the products appear $\mathbf{J}_M(\lambda_F)\hat{\mathbf
M}_1\sim\sin\chi$ and $\mathbf{J}_{sd}(\lambda_F)\hat{\mathbf
M}_1\sim\sin\chi$, which are small at $\chi\ll 1$ and at $|\pi-\chi|\ll
1$. We shall consider in this paper (see Section~\ref{section8}) only
small or close to $\pi$ angles $\chi$. It gives us the reason to neglect
further the products indicated that are proportional to $\sin\chi$.

The next step is to extract transverse part of the condition~\eqref{39}.
According to Eq.~\eqref{38}, this part for the forward current $j/e>0$ is
induced by the component
\begin{equation}\label{41}
  \mathbf J_\bot(\varepsilon)\equiv\left[\hat{\mathbf
  M}_2(\lambda_F),\,\left[\mathbf J(-\lambda_F),\,\hat{\mathbf M}_2(\lambda_F)\right]\right]
\end{equation}
and leads to the relation
\begin{equation}\label{42}
  \mathbf J_\bot(\varepsilon)=\mathbf J_M(\lambda_F)+\mathbf J_{sd}(\lambda_F).
\end{equation}

For backward current $j/e<0$ the other electron flux, namely,
$\mathbf{J}(\lambda_F)$, induces transverse (perpendicular to
$\mathbf{M}_1$) component while the flux $\mathbf{J}(\varepsilon)$ appears
purely longitudinal one being collinear with $\mathbf{M}_1$. This
interesrting process goes due to current dependent \emph{sd} exchange
effective field $\Delta\mathbf{H}_{sd}$ (see Eq.~\eqref{25}) produces
strong and localized influence on $\mathbf{M}_2$ by virtue the LLG
equation should be satisfied. The field $\Delta\mathbf{H}_{sd}$ is
directed parallel to $\mathbf{M}_1$ and therefore any transverse component
of the mobile electron magnetization $\Delta\mathbf{m}_2$ inside the SB
layer will decay and transfer to the lattice, the transformation being
analogous, in principal, to TST idea of the Refs.~\onlinecite{Slonczewski,
Berger}. Therefore, we may write the following induced transverse
component for backward current $j/e<0$:
\begin{equation}\label{43}
  \mathbf J_\bot(\lambda_F)=\left[\hat{\mathbf
  M}_1,\,\left[\mathbf J(\lambda_F),\,\hat{\mathbf
  M}_1\right]\right],
\end{equation}
which leads from Eq.~\eqref{39} to the relation
\begin{equation}\label{44}
  \mathbf J_\bot(\lambda_F)=-\mathbf J_M(\lambda_F)-\mathbf J_{sd}(\lambda_F).
\end{equation}

Transverse part of condition~\eqref{40} occurs the same for both forward
and backward current directions, $j/e>0$ and $j/e<0$, namely
\begin{equation}\label{45}
  \mathbf J_M(L-\varepsilon)=0.
\end{equation}

Now, we will try to represent conditions~\eqref{42} and~\eqref{44} in the
forms, which are more convenient for further calculations. Let us multiply
Eqs.~\eqref{42} and ~\eqref{44} by $\hat{\mathbf M}_2(\lambda_F)$
vectorially and use relations~\eqref{30},
~\eqref{32},~\eqref{41},~\eqref{43} and~\eqref{12}. Then, we get for
$j/e>0$
\begin{eqnarray}\label{46}
  &&\frac{\partial\hat{\mathbf M}_2}{\partial
  x}\Bigl|_{x=\lambda_F}=k\cdot\left[\hat{\mathbf M}_1,\,\hat{\mathbf
  M}_2(\lambda_F)\right]\nonumber \\
  &&-p\cdot\left[\hat{\mathbf
  M}_2(\lambda_F),\,\left[\hat{\mathbf M}_1,\, \hat{\mathbf
  M}_2(\lambda_F)\right]\right],
\end{eqnarray}
where two new parameters $k$ and $p$ appear, which are given by formulae
\begin{equation}\label{47}
  k=\frac{1}{aM_2}\left(\frac{\mu_B}{e}Q_1j-\tilde D_1\frac{\partial\Delta
  m_1}{\partial x}\Bigl|_{x=-\lambda_F}\right)\quad\text{for}\; j/e>0,
\end{equation}
\begin{equation}\label{48}
  k=\frac{1}{aM_2}\left(\frac{\mu_B}{e}Q_2j-\tilde D_2\frac{\partial\Delta
  m_2}{\partial x}\Bigl|_{x=\lambda_F}\right)\quad\text{for}\; j/e<0
\end{equation}
and
\begin{equation}\label{49}
  p=\frac{l_2h_{sd}}{AM_2},
\end{equation}
where $h_{sd}$ should be taken from formula~\eqref{26} for $j/e>0$ and
from~\eqref{27} for $j/e<0$.

Parameter $k$ originates from the contribution of mobile electron spin
fluxes $\mathbf J_{1\bot}(\varepsilon)$~\eqref{41} and $\mathbf
J_\bot(\lambda_F)$~\eqref{43}, and therefore it describes the spin
transfer torque action. Parameter $p$ originates from $\mathbf
J_{sd}(\lambda_F)$ flux~\eqref{32} and describes action of the spin
injection effective field. Let us substitute the before found expression
$\Delta m_1$ (see Eqs.~\eqref{19}--\eqref{22}) for $j/e>0$ into~\eqref{47}
and corresponding expression $\Delta m_2$ for $j/e<0$ into~\eqref{48} to
simplify both the expressions for $k$. It gives the following formulae
simplified for the thin layer \textbf{2} that is for $\lambda\ll 1$:
\begin{equation}\label{50}
  k=\frac{\mu_BQ_1j}{eaM_2}\frac{\nu}{\nu
  +\left(\hat{\mathbf M}_1\hat{\mathbf
  M}_2(\lambda_F)\right)^2}\quad\text{for}\;j/e>0
\end{equation}
and
\begin{equation}\label{51}
  k=\frac{\mu_BQ_1j}{eaM_2}\frac{\nu\left(\hat{\mathbf M}_1\hat{\mathbf
  M}_2(\lambda_F)\right)^2}{1
  +\nu\left(\hat{\mathbf M}_1\hat{\mathbf
  M}_2(\lambda_F)\right)^2}\quad\text{for}\;j/e<0.
\end{equation}

It is interesting to return now to the case of completely pinned layer
\textbf{1} (that is not only the lattice magnetization but mobile electron
spins also are pinned). We see from the first expression~\eqref{47} the
effective diffusion coefficient should be taken as $\tilde D_1\rightarrow
0$. Then it follows from the parameters $\nu_1$ and $\nu$ definitions made
after the formulae~\eqref{20},~\eqref{21} and~\eqref{26}:
$\nu_1\rightarrow\infty$, $\nu\equiv\nu_1\nu_2\rightarrow\infty$.
Substituting these limiting values to expressions~\eqref{50},~\eqref{51}
and to ~\eqref{26},~\eqref{27} we obtain
\begin{equation}\label{52}
    k=\frac{\mu_BQ_1}{aM_2}\frac{j}{e},\quad
    p=\frac{\mu_B\gamma\alpha_2Q_1\tau_2\lambda}{a\nu_2}\left|\frac{j}{e}\right|
\end{equation}
for a completely pinned layer \textbf{1} and for \underline{any sign of
the current} $j/e$. Therefore, the sign of the parameter $k$ does depend
on the sign of the current whereas parameter $p$ is positive at any sign
of the current. We shall see a little later this feature conserves in
general case of non-pinned mobile electron spins (see formulae~\eqref{59}
and \eqref{60}) and it plays significant role in determining the instability
character.

It remains now to rewrite the boundary condition~\eqref{45} for the second
interface $x=L$ in a more explicit form. Using the definition~\eqref{30}
and the equality $\left(\hat{\mathbf M}_2\cdot\partial\mathbf M_2/\partial
x\right)_{x=L-\varepsilon}\equiv 0$, we find
\begin{equation}\label{53}
  \frac{\partial\mathbf M_2}{\partial x}\Bigl|_{x=L-\varepsilon}=0.
\end{equation}

\section{Static states}\label{section7}
We start now to find solutions of LLG equation~\eqref{1} satisfying our
boundary conditions~\eqref{46} and~\eqref{53}. We confine ourselves
further to the situation when the current density $j$ is considered as a
fixed parameter given by an external circuit (the so called ``current
controlled regime''). Moreover, we assume an external field $\mathbf H$ is
applied along the positive direction of $z$ axis (see Fig.~\ref{fig1}),
and the anisotropy field $\mathbf{H}_a$ is parallel to that axis too
(remember: $H\le H_a\ll 4\pi M$). Then in absence of current, when there
is no coupling between the layers in our model, the layer \textbf{2}
magnetization $\hat{\mathbf M}_2$ is aligned along $z$ axis. Remember the
vectors $\hat{\mathbf M}_1$ and $\hat{\mathbf z}$ make, in general, an
angle $\chi$ in the model. Therefore we may write: $\hat{\mathbf
M}_1=-\hat{\mathbf y}\sin\chi+\hat{\mathbf z}\cos\chi$, where
$\hat{\mathbf x}$, $\hat{\mathbf y}$ and $\hat{\mathbf z}$ are basis
vectors of our coordinate system. After the current is turned on, the
direction of the vector $\hat{\mathbf M}_2$ may change. Our task now is to
search the vector $\hat{\mathbf M}_2$ change due to the current.

First, consider the static states possible when the fluctuations are
neglected but the current presents, $j/e\ne 0$. We denote the static state
magnetization as $\hat{\bar{\mathbf M}}_2$. To find the magnetization we
would solve the equation $\left[\hat{\bar{\mathbf
M}}_2,\,\mathbf{H}_{eff}\right]=0$, which follows from Eq.~\eqref{1}. The
last equation and the boundary conditions are essentially nonlinear and
nonuniform. Therefore general problem appears rather complicated even for
finding the static states.

Fortunately, there exists one simple situation, which is at the same time
the most actual one, namely, the situation with the angle $\chi=0$ or
$\chi=\pi$. In the situation all the boundary conditions and the equation
of motion for $\hat{\bar{\mathbf M}}_2$ are satisfied for
\begin{equation}\label{54}
  \hat{\bar M}_{2x}=\hat{\bar M}_{2y}=0,\quad\hat{\bar M}_{2z}=\pm 1.
\end{equation}

The solution written can be easy justified by substitution of`\eqref{54}
into the equation and boundary conditions. We consider further only these
last indicated values of the angle $\chi$ and therefore investigate the
stability only for the static states~\eqref{54}. For definiteness, we will
consider only the angle $\chi=\pi$ and therefore have
\begin{equation}\label{55}
  \hat{\bar{\mathbf M}}_1=-\hat{\mathbf z}.
\end{equation}

As it is seen from Eqs.~\eqref{54} and~\eqref{55}, two situations are
possible: either vectors $\hat{\bar{\mathbf M}}_2$ and $\hat{\bar{\mathbf
M}}_1$ are parallel each other (\textbf{P} state) or they are antiparallel
(\textbf{AP} state). We shall consider further the both possibilities.

\section{Fluctuations. Dispersion relation}\label{section8}
Passing to consideration of fluctuations $\Delta\hat{\mathbf M}_2$, we
introduce them by the following relation:
\begin{equation}\label{56}
    \hat{\mathbf M}_2=\pm\hat{\mathbf z}+\Delta\hat{\mathbf M}_2,
\end{equation}
where $\Delta\hat{M}_2\equiv\left|\Delta\hat{\mathbf M}_2\right|\ll 1$. We
linearize the LLG equation with respect to $\Delta\hat{\mathbf M}_2$ and,
according to~\eqref{25}, lay $\Delta\mathbf{H}_{sd}\equiv0$ outside the SB
layer, that is for $x>\lambda_F$. The demagnetization field takes the form
$\mathbf{H}_d=-4\pi M_2\Delta\hat{M}_{2x}\hat{\mathbf x}$. Then the
equations for fluctuations $\Delta\hat{M}_{2x,y}\sim\exp(-i\omega t)$
reduce to
\begin{eqnarray}\label{57}
\frac{\partial^2\Delta\hat{M}_{2x}}{\partial
x^2}-\frac{\Omega_1}{a}\Delta\hat{M}_{2x}-\frac{i\omega\hat{\bar
M}_{2z}}{a}\Delta\hat{M}_{2y}=0,\nonumber \\
\frac{\partial^2\Delta\hat{M}_{2y}}{\partial
x^2}-\frac{\Omega_2}{a}\Delta\hat{M}_{2y}+\frac{i\omega\hat{\bar
M}_{2z}}{a}\Delta\hat{M}_{2x}=0,
\end{eqnarray}
where two characteristic frequencies are introduced:
\begin{eqnarray}\label{58}
  \Omega_1&=&\gamma\left(H\hat{\bar
  M}_{2z}+H_a+4\pi M_2\right)-i\kappa\omega,\nonumber \\
  \Omega_2&=&\gamma\left(H\hat{\bar
  M}_{2z}+H_a\right)-i\kappa\omega.
\end{eqnarray}

We need now to linearize boundary conditions and use them together with
equations~\eqref{57}. Parameters $k$ and $p$ \eqref{47}--\eqref{49} should
be taken for the static state chosen in the linear approximation. It means
we need to rewrite general expressions~\eqref{49}--\eqref{51} in a more
specific form, which appears valid for any sign of current:
\begin{equation}\label{59}
  k=\frac{\mu_BQ_1}{aM_2}\frac{\nu}{1+\nu}\cdot\frac{j}{e},
\end{equation}
\begin{equation}\label{60}
  p=\frac{\mu_B\gamma\alpha_2\tau_2Q_1}{a}\frac{\left[\lambda\nu_1(\nu-1)-2b\nu\hat{\bar
  M}_{2z}\right]}{(1+\nu)^2}\left|\frac{j}{e}\right|.
\end{equation}

Linearized boundary conditions after the transformations mentioned acquire
the following form:
\begin{equation}\label{61}
  \frac{\partial\Delta\hat{M}_{2x}}{\partial
  x}\biggr|_{x=\lambda_F}=k\Delta\hat{M}_{2y}(\lambda_F)-p\hat{\bar
  M}_{2z}\Delta\hat{M}_{2x}(\lambda_F),
\end{equation}
\begin{equation}\label{62}
  \frac{\partial\Delta\hat{M}_{2y}}{\partial
  x}\biggr|_{x=\lambda_F}=-k\Delta\hat{M}_{2x}(\lambda_F)-p\hat{\bar
  M}_{2z}\Delta\hat{M}_{2y}(\lambda_F),
\end{equation}
\begin{equation}\label{63}
  \frac{\partial\Delta\hat{M}_{2x}}{\partial
  x}\biggr|_{x=L-\varepsilon}=\frac{\partial\Delta\hat{M}_{2y}}{\partial
  x}\biggr|_{x=L-\varepsilon}=0.
\end{equation}

Let us substitute $\Delta\hat{M}_{2x,y}\sim\exp(iqx')$ into
equations~\eqref{57} and define $x'=x-\lambda_F$. Then we find four
possible characteristic wave numbers:
\begin{equation}\label{64}
  q_\pm^2=-\frac{1}{2a}\left[\Omega_1+\Omega_2\pm\sqrt{(\Omega_1-\Omega_2)^2+4\omega^2}\right].
\end{equation}

Then, the general solution of our linearized equations~\eqref{57} may be
presented as follows
\begin{eqnarray}\label{65}
    \Delta\hat{M}_{2x}(x')=A\cos q_-x'+B\sin q_-x'\nonumber \\
    +C\cos q_+x'+D\sin q_+x',\nonumber \\
    \Delta\hat{M}_{2y}(x')=\xi(A\cos q_-x'+B\sin q_-x')\nonumber \\
    +\xi^{-1}(C\cos q_+x'+D\sin q_+x'),
\end{eqnarray}
where we took $\mathrm{Re}\,q_\pm>0$ and introduced a ``distribution
coefficient''
\begin{equation}\label{66}
  \xi=2i\omega\hat{\bar M}_{2z}\left[\Omega_2
  -\Omega_1+\sqrt{(\Omega_1-\Omega_2)^2+4\omega^2}\right]^{-1}.
\end{equation}

Constants $A,\,B,\,C,\,D$ should be found now from the boundary
conditions~\eqref{61}--\eqref{63}. Substituting~\eqref{66} into the
conditions and equating the determinant to zero, we find the dispersion
relation for the fluctuations in the most general form:
\begin{eqnarray}\label{67}
  \left(q_-\tan q_-L+p\hat{\bar M}_{2z}\right)\left(q_+\tan q_+L+p\hat{\bar
  M}_{2z}\right)\nonumber \\
  +k^2+\frac{2\xi k}{1-\xi^2}\left(q_-\tan q_-L-q_+\tan q_+L\right)=0.
\end{eqnarray}

We may compare Eq.~\eqref{67} with before derived dispersion relation in
Refs.~\onlinecite{Gulyaev4,Gulyaev5}. Note, parameter $\xi^2\ne1$. It can
be seen directly from the definitions~\eqref{66} and~\eqref{58}. Taking
this into account, we see the relations mentioned may be exactly reduced
to each other but for the case $k=0$ only, that is for the case of the LSI
effective field mechanism is dominated. It appears due to rather crude
approximation used in Refs.~\onlinecite{Gulyaev4,Gulyaev5} when deriving
the dispersion relation. Contrary to this, our last expression~\eqref{67}
represents more correct and complete dispersion relation, which describes
joint action of TST torque and LSU effective field mechanisms.

\section{Instability of fluctuations}\label{section9}
We need further to solve the dispersion relation~\eqref{67} and find
eigenfrequencies $\omega$ for fluctuations. In particular, it allows us to
determine the current driven instability condition, that is the condition
when $\mathrm{Im}\,\omega>0$. Generally speaking, the transcendental
equation~\eqref{67} may be solved numerically for a number of
experimentally given junction parameters. However, the actual task now is
to develop a general representation about the switching process rather
than to get specific numerical results. To fulfil the task, we may try to
solve the equation analytically but only for a special case of nearly
uniform fluctuations which satisfy the equality $\tan q_-L\approx q_-L$.

It is convenient to use Eqs.~\eqref{64},~\eqref{66} and rewrite
Eq.~\eqref{67} in the form
\begin{eqnarray}\label{68}
&&\Omega_1+\Omega_2-\sqrt{(\Omega_1-\Omega_2)^2+4\omega^2}\nonumber \\
&&=\left(\frac{2a}{L}\right)\Biggl[p\hat{\bar M}_{2z}+\frac{k^2}{q_+\tan
q_+L+p\hat{\bar M}_{2z}}\nonumber \\
&&+\frac{2i\omega k\hat{\bar
M}_{2z}}{a\left(q_+^2-q_-^2\right)}\left(1-\frac{q_-^2L+p\hat{\bar
M}_{2z}}{q_+\tan q_+L+p\hat{\bar M}_{2z}}\right)\Biggr].
\end{eqnarray}

If current is absent (that is $k=p=0$), we obtain from Eq.~\eqref{68}
\begin{eqnarray}\label{69}
  \frac{\omega}{\omega_0}&=&\sqrt{(1+\kappa^2)\left(1
  +\frac{h\hat{\bar
  M}_{2z}}{h_a}\right)-\frac{\kappa^2}{4h_a}}-i\frac{\kappa}{2\sqrt{h_a}},\nonumber
  \\ \omega_0&=&\frac{4\pi\gamma M_2}{1+\kappa^2}\sqrt{h_a},
\end{eqnarray}
where we remain only positive frequency $\mathrm{Re}\,\omega\ge0$ and
introduced dimensionless fields: $h=H/4\pi M_2$ and $h_a=H_a/4\pi M_2$.
According to typical values of parameters, we have $H\sim10$--100 Oe,
$H_a\sim100$ Oe, $M_2\sim10^3$ G and therefore $h,\,h_a\lesssim0.01$. We
neglect small corrections originated due to higher powers of $h,\,h_a$ in
Eq.~\eqref{69}. As it is seen from Eq.~\eqref{69}, the fluctuations
without current are, of course, stable due to the Gilbert damping.

Return now to the situation when the current is turned on ($j\ne0$) but is
not too large in magnitude to allow preserve in Eq.~\eqref{68} only linear
in current summands. The justification of the supposition will be
considered later, at the end of the section. If we accept now that the
supposition is valid, it becomes possible to insert $q_-=0$ into the
square brackets of Eq.~\eqref{68}. It means, in particular, we may lay
$\sqrt{(\Omega_1-\Omega_2)^2+4\omega^2}=\Omega_1+\Omega_2$ in the
definition~\eqref{64} of $q_+$ and therefore use the expression
\begin{equation}\label{70}
  q_+\approx i\sqrt{\frac{\Omega_1+\Omega_2}{a}}
\end{equation}
to substitute it to the right hand side of Eq.~\eqref{68}. Moreover, we
should neglect summands proportional to $k^2$ and $kp$ too. Formally,
these simplifications become valid if the following conditions are
fulfilled:
\begin{equation}\label{71}
  \frac{a|p|}{L\left|\Omega_1+\Omega_2\right|}\ll
  1,\quad\left|\frac{k^2}{pq_+}\right|\ll 1,\quad\left|\frac{p}{q_+}\right|\ll
  1.
\end{equation}

Then we resolve Eq.~\eqref{68} with respect to square root and square the
result. Some summands appear containing second (or higher) power of the
current ($k^2,\,kp$ and so on). These summands are small also and should
be neglected due to the conditions~\eqref{71}. After we make the
simplifications and substitute~\eqref{70}, we obtain the following square
equation for the eigenfrequency $\omega$:
\begin{eqnarray}\label{72}
    &&\omega^2(1+\kappa^2)+i\kappa\omega\left[\mathrm{Re}\,(\Omega_1+\Omega_2)-\frac{2ap\hat{\bar
    M}_{2z}}{L}-\frac{2ak\hat{\bar
    M}_{2z}}{\kappa L}\right]\nonumber \\
    &&-\left[\mathrm{Re}\,(\Omega_1\Omega_2)-\frac{ap\hat{\bar
    M}_{2z}}{L}\mathrm{Re}\,(\Omega_1+\Omega_2)\right]=0.
\end{eqnarray}

We should solve the equation~\eqref{72} using the definitions
of the parameters $k$ and $p$ given by Eqs.~\eqref{59} and~\eqref{60}.
Then we obtain finally
\begin{eqnarray}\label{73}
  &&\frac{\omega}{\omega_0}=-i\frac{\kappa_0}{2\sqrt{h_a}}\left[\frac{\kappa}{\kappa_0}
  -\frac{\hat{\bar M}_{2z}(j/e)}{(j_0/|e|)}\right]\nonumber \\
  &&+\Biggl\{(1+\kappa^2)\left(1+\frac{\hat{\bar M}_{2z}h}{h_a}
  -\frac{\hat{\bar M}_{2z}|j|}{j_0}\right)\nonumber \\
  &&-\left(\frac{\kappa_0}{2\sqrt{h_a}}\right)^2\left[\frac{\kappa}{\kappa_0}
  -\frac{\hat{\bar M}_{2z}(j/e)}{(j_0/|e|)}\right]^2\Biggr\}^{1/2},
\end{eqnarray}
where we used the following simplifying assumption
\begin{equation}\label{74}
  \eta\equiv\frac{|k|}{\kappa|p|}\gg 1,
\end{equation}
which is really true for many actual cases of small enough Gilbert
parameter $\kappa$. Formula~\eqref{73} reduces to~\eqref{69} in the limit
$j\rightarrow0$. The following additional system characteristics are
introduced in Eq.~\eqref{73}:

\underline{characteristic damping}
\begin{equation}\label{75}
  \kappa_0=\frac{2h_a\nu(\nu+1)}{\gamma\alpha_2\tau_2M_2\left[\lambda\nu_1(\nu-1)
  -2b\nu\hat{\bar M}_{2z}\right]}
\end{equation}
and

\underline{characteristic current density}
\begin{equation}\label{76}
  \left(j_0/|e|\right)=\frac{4\pi M_2l_2h_a\lambda(\nu+1)^2}{\mu_B\alpha_2\tau_2Q_1
  \left[\lambda\nu_1(\nu-1)
  -2b\nu\hat{\bar M}_{2z}\right]}.
\end{equation}
Here the matching parameters $\nu_1$ and $\nu$ are exactly the same as
they were defined after formulae~\eqref{20},~\eqref{21} and~\eqref{26}. We
may substitute the following typical values: $\nu_2\sim1$, $\nu_1\gg 1$
and $\nu\gg 1$, $\lambda\sim 0.1$, $l\sim 17\,\rm nm$, $\alpha_2\sim
2\times 10^4$, $\tau_2\sim 3\times 10^{-13}\,\rm s$, $Q_1\sim Q_2 \sim
0.3$. Then we get the estimations: $|\kappa_0|\sim3\times 10^{-2}$,
$|j_0|\sim 1\times 10^7\,\rm A/cm^2$.

Expression~\eqref{73} allows write the following two conditions of
instability ($\mathrm{Im}\,\omega >0$):
\begin{equation}\label{77}
    \hat{\bar M}_{2z}\frac{|j|}{j_0}>1+\frac{\hat{\bar M}_{2z}h}{h_a},
\end{equation}
\begin{equation}\label{78}
  \hat{\bar M}_{2z}\frac{j/e}{j_0/|e|}>\frac{\kappa}{\kappa_0}.
\end{equation}

The fulfillment of any of the conditions~\eqref{77} or~\eqref{78} or both
of them together is sufficient for instability. The condition~\eqref{77}
originates from $p$ parameter and therefore refers to LSI mechanism. The
condition~\eqref{78} originates from $k$ parameter and refers to TST
mechanism. The threshold current $j_{th}$ has a minimal module, which is
necessary to reach the instability according to Eqs.~\eqref{77}
and~\eqref{78}, that is
\begin{equation}\label{79}
  j_{th}=|j_0|\cdot\min\left(\left|1+\frac{\hat{\bar
  M}_{2z}h}{h_a}\right|,\;\frac{\kappa}{|\kappa_0|}\right).
\end{equation}

As it is seen, the threshold for LSI mechanism does not depend on the
Gilbert constant $\kappa$, while the TST threshold is directly
proportional to it. Because of experimentally revealed damping at room
temperature~\cite{Ingvarsson} $\kappa\approx(2\div5)\times10^{-2}$
corresponds to the above given estimation
$\kappa\sim|\kappa_0|\sim3\times10^{-2}$, we may expect any of the
mechanism, in principle, is responsible for the instability. Note, the
parameters $j_0$ and $\kappa_0$, according to the definitions~\eqref{75}
and~\eqref{76}, may change their signs and become negative. But it may
occur only simultaneously for $j_0$ and $\kappa_0$. Therefore, the TST
condition~\eqref{78} does not depend on the sign at all, but to provide
the LSI instability condition be fulfilled we should change the sign of
$\hat{\bar{M}}_{2z}$ only.

Let us take for further estimations:
\begin{equation}\label{80}
  \left|1+\frac{\hat{\bar M}_{2z}h}{h_a}\right|\sim 1\quad\frac{\kappa}{|\kappa_0|}\le
  1\quad\text{or}\quad\frac{\kappa}{|\kappa_0|}\sim 1
\end{equation}
and moreover $\nu\gg 1$, $\nu_2\sim 1$ and all the estimations from the
text are valid. Then, we substitute these parameters into
conditions~\eqref{71} and get
\begin{equation*}
  \frac{a|p|}{L|\Omega_1+\Omega_2|}\sim 10^{-2},\;
  \left|\frac{k^2}{pq_+}\right|\sim
  10^{-4},\;\left|\frac{p}{q_+}\right|\sim 10^{-2}.
\end{equation*}

We see, therefore, the conditions~\eqref{71} are good satisfied for the
threshold current and even above it. The last estimations justify our
above-stated analytic approach to the solution of the dispersion
relation~\eqref{67}. It is interesting to stress, our eigenfrequency
$\omega$ \eqref{73} appears essentially nonlinear function of the current
$j/j_0$ in spite of the linearization conditions~\eqref{71}.

\section{State diagram and hysteresis}\label{section10}
\begin{figure}
\includegraphics[width=8.6cm]{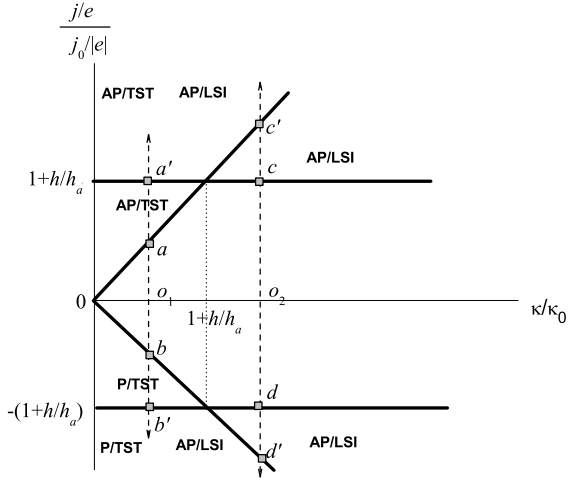}
\caption{State diagram. Threshold current versus damping (qualitatively)
at $\nu_1(\nu-1)\lambda\gg 2b\nu$, $\nu_1\gg 1$, $\nu_2=1$. Heavy lines
show boundaries of the regions. The following
notations are introduced: \textbf{AP/LSI} means that \textbf{AP} initial
state is unstable due to LSI mechanism; \textbf{P/TST} means that
\textbf{P} initial state is unstable due to TST mechanism, and so on. The
region around the line $j=0$ without any designations corresponds to
stability. Points of switching: \textit{a}~--- switching
$\mathbf{AP\rightarrow P}$, \textit{b}~--- switching $\mathbf{P\rightarrow
AP}$, \textit{c} and \textit{d}~--- switching $\mathbf{AP\rightarrow P}$.
Points of oscillations: $b'$ and $d'$.} \label{fig3}
\end{figure}

We discuss now in detail the instability conditions~\eqref{77}
and~\eqref{78}. Let us draw the corresponding threshold curves in the
current to damping plane~--- the ``state diagram''. The result is
presented qualitatively in Fig.~\ref{fig3}. We confined ourselves only the
case of positive parameters $j_0>0$ and $\kappa_0>0$ and built up the
boundaries of the regions indicated. The condition~\eqref{77} can be
satisfied for $\hat{\bar M}_{2z}=1$ only. It means the LSI mechanism can
excite the instability of \textbf{AP} state only to switch it into
\textbf{P} state. That is true whatever sign is taken for the current.
Another situation appears in connection with the condition~\eqref{78}. For
forward current $j/e>0$ the only value $\hat{\bar M}_{2z}=1$ is
acceptable, but for backward current $j/e<0$ it is necessary to have
$\hat{\bar M}_{2z}=-1$. It means the TST mechanism excites the instability
of different states depending on the sign of the current: forward current
excites \textbf{AP} state, but backward current excites \textbf{P} state.
All these possibilities are seen from Fig.~\ref{fig3}.

\begin{figure}
\includegraphics[width=8.6cm]{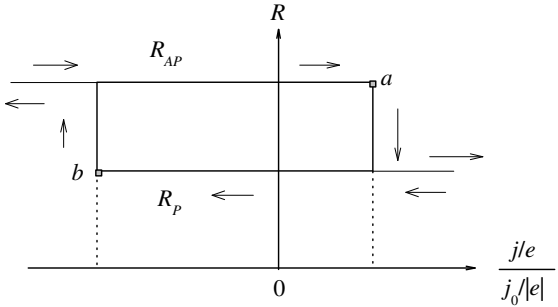}
\caption{Hysteretic behavior of junction resistance $R$: $R_{AP}$, $R_P$
are resistances of \textbf{AP} and \textbf{P} states,
respectively.}\label{fig4}
\end{figure}

Two notes should be made here. First, let us go along the line $o_1a$
having \textbf{AP} state. In the point $a$ the state becomes unstable and
will be switched to the stable \textbf{P} state. We may return then to the
point $b$, where the \textbf{P} state becomes unstable and will be
switched to the initial \textbf{AP} state, which now is stable, and so on.
Any switching process leads to change of the junction resistance, the
higher resistance being correspond to \textbf{AP} state and the lower
resistance to \textbf{P} state. Such a behavior is illustrated in
Fig.~\ref{fig4} and similar observations were revealed experimentally in
many papers.

The second note concerns the intersections of our line $o_1a$ in
Fig.~\ref{fig3} or some other lines ($o_2c$, for example) with the
threshold curves in the points $b',\,d'$ and so on. In the
points all the initial states, \textbf{AP} and \textbf{P}, become
unstable. Therefore, no switching can appear and instead some new time
dependent oscillatory state, apparently, should form. This is a very
interesting possibility, which correlates with the well known experimental
observations of a microwave emission from the junctions above the
instability threshold (see, e.g., Ref.~\onlinecite{Tsoi}). We plan to
investigate the possibility in more detail elsewhere.

\section{Spectrum softening and increment}\label{section11}
\begin{figure}
\includegraphics[width=8.6cm]{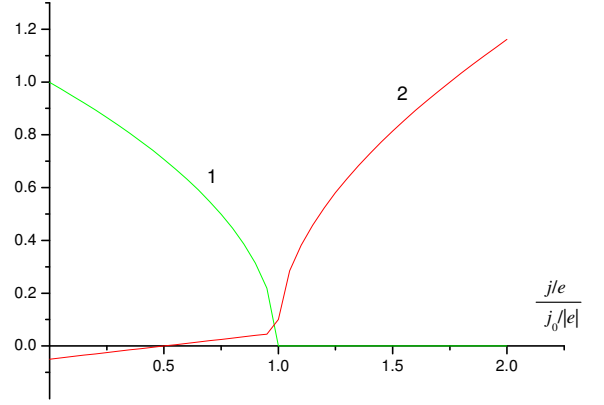}
\caption{(Color online) Current dependence of eigenfrequency
$\mathrm{Re}\,\omega/\omega_0$ (curve 1) and decrement
$\mathrm{Im}\,\omega/\omega_0$ (curve 2). Damping constant
$\kappa/\kappa_0=0.5$.}\label{fig5}
\end{figure}

\begin{figure}
\includegraphics[width=8.6cm]{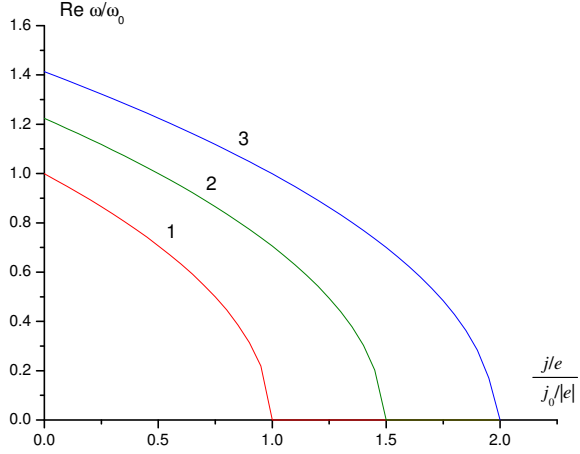}
\caption{(Color online) Softening of eigenfrequency for different external
fields: 1~--- $h/h_a=0$, 2~--- $h/h_a=0.5$, 3~--- $h/h_a=1.0$. Parameters:
$\nu\gg 1$, $b\sim 1$, $\kappa_0/(2\sqrt{h_a})=0.1$.}\label{fig6}
\end{figure}

\begin{figure}
\includegraphics[width=8.6cm]{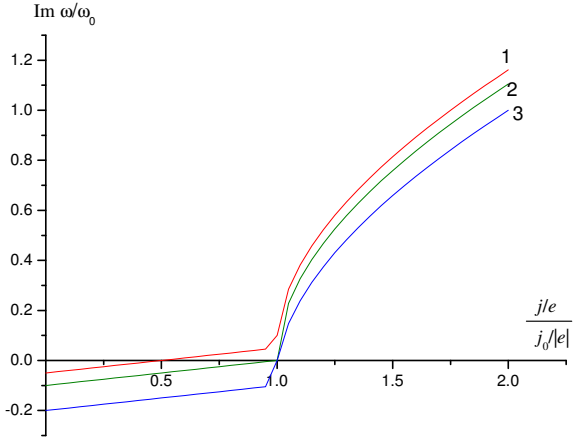}
\caption{(Color online) Increments for different damping: 1~---
$\kappa/\kappa_0=0.5$, 2~--- $\kappa/\kappa_0=1.0$, 3~---
$\kappa/\kappa_0=2.0$. Parameters: $\nu\gg 1$, $b\sim 1$, $h=0$,
$\kappa_0/(2\sqrt{h_a})=0.1$.}\label{fig7}
\end{figure}

We discuss now he question: how spectrum ($\mathrm{Re}\,\omega$) and
increment ($\mathrm{Im}\,\omega$) behave when the current comes near the
threshold of instability. To answer, we consider Eq.~\eqref{73} for
eigenfrequency and calculate numerically the quantities of interest for
some typical example. Figs.~\ref{fig5}--\ref{fig7} demonstrate the most
significant results.

According to Fig.~\ref{fig5}, a normalized spectrum
$\mathrm{Re}\,\omega/\omega_0$ falls to zero, that is softens, when the
current rises nearing to the LSI threshold. As it was remarked in
Ref.~\onlinecite{Gulyaev4}, the softening appears due to LSI driven
reorientation phase transition at the threshold. Indeed, such a softening
is the general property of similar phase transitions. Contrary to this,
the TST has no relation to any phase transition~\cite{Gulyaev4} and is
simply a mechanism to deliver energy into the magnetic lattice.
correspondingly, increment $\mathrm{Im}\,\omega/\omega_0$ changes its sign
and becomes positive not at the LSI threshold but a little below if
damping is small enough, namely, for $\kappa/\kappa_0=0.5$ in our example.
It is significant to stress the increment rises faster when current
increases in the region where the LSI mechanism dominates. If we estimate
$\omega_0\sim10^{10}\,\rm s^{-1}$, then the increment may rise from
approximately $\mathrm{Im}\,\omega\sim10^{9}\,\rm s^{-1}$ for TST
mechanism dominated to $\mathrm{Im}\,\omega\sim10^{10}\,\rm s^{-1}$ for
LSI mechanism dominated.

Fig.~\ref{fig6} shows the spectrum for different values of an external
magnetic field. Eigenfrequencies rise with increasing the field. The LSI
threshold shifts toward large currents. But the fact of spectrum
softening, of course, remains.

Fig.~\ref{fig7} shows the increment for different values of a damping
constant and in absence of external magnetic field. As the damping
increases, the increment at a given current decreases for any mechanisms
(TST and LSI) but LSI threshold is fixed, that is it does not depend on
the damping. The TST threshold rises with damping and reaches the LSI
threshold at the end.

\section{Conclusion}\label{section12}
A phenomenological approach is developed in the theory of spin-valve type
ferromagnetic junctions. New vector boundary conditions are derived, which
represent the continuity conditions for different spin fluxes:
longitudinal spin flux of mobile electrons and transverse spin fluxes of
the magnetic lattice including the so called ``effective field transverse
spin flux''. Spin transfer torque is considered as a boundary condition.
It describes the result of abrupt transformation of the mobile electron
transverse spin flux to the spin fluxes of the lattice. Joint action of
two electric current effects is investigated: the nonequilibrium
longitudinal spin injection effective field and the transverse spin
transfer surface torque. General macroscopic dynamic equations are
linearized for fluctuations and solved using the boundary conditions
mentioned. Dispersion relation is derived and solved.

The calculation model is extended significantly to make it adequate the
experimental situation as far as it is possible. In particular, the mobile
electrons are considered as having non-pinned spins in all the contacting
layers. Lattice magnetization is assumed completely pinned in one of the
magnetic layers and free in the other magnetic layer. Theory is developed
for any current direction in the spin-valve junction (forward or
backward).

Some critical value $\kappa_0$ of the well known Gilbert damping constant
$\kappa$ is introduced for the first time. Spin transfer torque gives the
dominant contribution to instability threshold for small enough constants
$\kappa<\kappa_0$, while the spin injection contribution dominates for
$\kappa>\kappa_0$. Typical estimation gives $\kappa_0\sim3\times10^{-2}$
and it approximately corresponds to the experimentally found Gilbert
constant values for ferromagnetic metals at room temperature. It means
both the two mechanisms may be responsible for the measured thresholds.

The mechanism of magnetic switching by the backward current and hysteretic
behavior is proposed for the first time. As it appears, a fine interplay
of spin transfer torque and spin injection is necessary to provide the
hysteresis. The peculiarity of the process may be understood as follows.
Spin injection effective field is localized in the free layer near the
boundary between pinned and free layers. Spins of mobile electrons in
the backward current have a component transverse to the effective
field and therefore perform a precession in the field. This precession
leads to torque generation and switching in complete accordance with the
transverse spin transfer mechanism. The consequent realization of forward
and such type of backward switching leads to hysteretic behavior shown in
Fig.~\ref{fig4}. The existence of the hysteresis is consistent with
experimental data.

The state diagram was built up that divided the current--damping plane
(Fig.~\ref{fig3}) into regions of fluctuation stability and regions of
various type instabilities. For example, for forward currents the
\textbf{AP} initial states may become unstable and \textbf{P} states remain
stable. In the regions the switching processes (i.e.,
$\mathbf{AP}\rightarrow\mathbf{P}$) become possible. This diagram revealed
for the first time the existence of another type regions for backward
currents where \textbf{AP} and \textbf{P} initial states become unstable
simultaneously. This last type of regions may correspond to formation of
time dependent (oscillatory) states as a result of the instability
development.

The effect of softening was predicted for the layer spin wave
eigenfrequencies when current rises and comes to the instability threshold
(Figs.~\ref{fig5} and \ref{fig6}). This effect is interesting because it
distinguishes directly between TST and LSI mechanisms, since only the LSI
mechanism leads to the softening.

The estimations were made of instability increments as functions of the
current. It was shown for the first time that LSI mechanism leads to much
larger increments in comparison with TST mechanism. For example, if we
take spin wave eigenfrequency $\mathrm{Re}\,\omega\sim10^{10}\,\rm
s^{-1}$, then we get increment $\mathrm{Im}\,\omega\sim10^{9}\,\rm s^{-1}$
for TST mechanism dominated and $\mathrm{Im}\,\omega\sim10^{10}\,\rm
s^{-1}$ for LSI mechanism dominated. The increment rises for any mechanism
if current increases and damping decreases.

\begin{acknowledgments}
The authors are thankful to Professor Sir Roger Elliott for attraction of
their attention to the problem of ferromagnetic junction switching and for
fruitful collaboration.

The work was supported by Russian Foundation for Basic Research, Grant
No.~06-02-16197.
\end{acknowledgments}

\appendix*
\section{Transverse spin transfer torque considered as a localized or as a uniformly
spread}\label{appendix}

We try now to show the connections between two models discussed till now:
traditional model of uniformly spread TST torque and recently introduced
(Refs.~\onlinecite{Gulyaev4,Gulyaev5,Gulyaev7,Gulyaev9} and this paper)
localized TST torque.

We start from the linearized equations and boundary conditions for
fluctuations~\eqref{57} and~\eqref{61}--\eqref{63}. The boundary
conditions~\eqref{61} and~\eqref{62} may be inserted into Eq.~\eqref{57}
using a singular $\delta$-function, namely,
\begin{eqnarray}\label{A1}
    &&\frac{\partial^2\Delta\hat{M}_{2x}}{\partial
    x^2}-\frac{\Omega_1}{a}\Delta\hat{M}_{2x}
    -\frac{i\omega\hat{\bar{M}}_{2z}}{a}\Delta\hat{M}_{2y}\nonumber \\
    &&=\left[k\Delta\hat{M}_{2y}-p\hat{\bar{M}}_{2z}\Delta\hat{M}_{2x}\right]
    \delta(x-\lambda_F),\nonumber \\
    &&\frac{\partial^2\Delta\hat{M}_{2y}}{\partial
    x^2}-\frac{\Omega_2}{a}\Delta\hat{M}_{2y}
    +\frac{i\omega\hat{\bar{M}}_{2z}}{a}\Delta\hat{M}_{2x}\nonumber \\
    &&=\left[-k\Delta\hat{M}_{2x}-p\hat{\bar{M}}_{2z}\Delta\hat{M}_{2y}\right]
    \delta(x-\lambda_F).
\end{eqnarray}

To check the correspondence of the Eqs.~\eqref{A1}
and~\eqref{57},~\eqref{61},~\eqref{62}, let us integrate~\eqref{A1} over
the interval $[0,\,\lambda_F+0]$. Then, we have for the derivative
\begin{eqnarray}\label{A2}
  \int\limits_0^{\lambda_F}\frac{\partial^2\Delta\hat{M}_{2x}}{\partial
    x^2}\,dx&=&\left[\frac{\partial\Delta\hat{M}_{2x}}{\partial
    x}\Biggl|_{x=\lambda_F}-\frac{\partial\Delta\hat{M}_{2x}}{\partial
    x}\Biggl|_{x=0}\right]\nonumber \\
    &=&\frac{\partial\Delta\hat{M}_{2x}}{\partial
    x}\Biggl|_{x=\lambda_F}
\end{eqnarray}
and analogous expression for the other derivative, which may be got by a
replacement $x\rightarrow y$, where we accept free spins at the plane
$x=0$, that is
\begin{equation}\label{A3}
    \frac{\partial\Delta\hat{M}_{2x}}{\partial x}\Biggl|_{x=0}
    =\frac{\partial\Delta\hat{M}_{2y}}{\partial x}\Biggl|_{x=0}=0.
\end{equation}

Moreover, we should neglect all the terms, which are proportional to small
length $\lambda_F\rightarrow0$. Then we get immediately the boundary
conditions in the form~\eqref{61},~\eqref{62}. Therefore, two completely
equivalent approaches may be used: either we solve our previous
equations~\eqref{57} satisfying nonuniform system of boundary
conditions~\eqref{61}--\eqref{63}, or we solve the singular
equations~\eqref{A1} with uniform boundary conditions~\eqref{63}
and~\eqref{A3}.

The second approach appears convenient to pass to uniform model. To do
this, we should spread uniformly through the whole layer \textbf{2}
thickness $L$ the localized excitation in~\eqref{A1}, that is we should
take
\begin{equation}\label{A4}
  \delta(x-\lambda_F)\rightarrow\frac{1}{L}.
\end{equation}

After this is done, the equations~\eqref{A1} become spatially uniform and
exactly coincide with the linearized uniform model. The last statement may
be directly confirmed if we start from LLG equation with the well known
additional double vector product term introduced in the original
works~\cite{Slonczewski, Berger}. The boundary conditions~\eqref{63}
and~\eqref{A3} are uniform also. It shows we can find a uniform solution
for fluctuations. To simplify extremely the calculations, we suppose the
thickness of the layer is sufficiently small and spin injection may be
omitted. Then the following dispersion relation for eigenfrequency becomes
valid in linear in current approximation:
\begin{eqnarray}\label{A5}
  (1+\kappa^2)\omega^2&+&i\kappa\omega\left[\mathrm{Re}\,(\Omega_1+\Omega_2)
  -\frac{2ak}{\kappa L}\hat{\bar{M}}_{2z}\right]\nonumber \\
  &-&\mathrm{Re}\,(\Omega_1\Omega_2)=0.
\end{eqnarray}

This dispersion relation coincides with the corresponding
relation of our localized torque model~\eqref{72} taken at $p=0$. The
condition of instability that follows from ~\eqref{A5} may be written in
the form
\begin{equation}\label{A6}
  \hat{\bar{M}}_{2z}\cdot\frac{j_{th,\bot}}{e}>\frac{2\pi\gamma M^2_2\kappa
   L(1+\nu)}{\mu_B\nu Q_1}.
\end{equation}

As it is seen, forward current ($j_{th,\bot}/e>0$) leads to instability of
\textbf{AP} states only (because of to satisfy~\eqref{A6} we should take
$\hat{\bar{M}}_{2z}=1$) and, vise versa, backward current
($j_{th,\bot}/e<0$) leads to instability of \textbf{P} states (we should
take $\hat{\bar{M}}_{2z}=-1$). This conclusion is in agreement with our
localized torque theory (see Section~\ref{section10} of this paper).
Numerical estimation gives $\left|j_{th,\bot}/e\right|\sim10^7\,\rm
A/cm^2$.

However, we should stress here the correct understanding of the problem
cannot be possible without including both the mechanisms (TST and LSI)
acting simultaneously. First, the LSI was included implicitly in the
condition~\eqref{A6} for backward current switching. Really, the existence
of \emph{sd} exchange effective field
$\Delta\mathbf{H}_{sd}\sim\delta(x-\lambda_F)$ (see Eq.~\eqref{25}) is
necessary to understand the backward transverse spin flux appearing in the
layer \textbf{2} and transformation of this flux into the lattice flux.
The relation~\eqref{44} describes such a transformation and allows
calculate the TST parameter $k$~\eqref{52}.

Moreover, the results of the paper show the LSI mechanism manifests itself
significantly and in different ways. For example, the instability
threshold does not rise with increasing the Gilbert damping constant
$\kappa$ but reaches some limiting value due to LSI mechanism (see Fig.
~\ref{fig3}). Increment of instability may be more than order of magnitude
larger due to LSI (see Fig.~\ref{fig5}). Interplay of TST and LSI
mechanisms makes it possible to provide the instability simultaneously of
\textbf{P} and \textbf{AP} states (Fig.~\ref{fig3}). It means, no
switching but rather new time dependent final nonlinear state may appear
as a result of the instability development.

\end{document}